# Building Smart Communities with Cyber-Physical Systems

**Feng Xia**
School of Software
Dalian University of Technology
Dalian 116620, China
f.xia@acm.org

**Jianhua Ma**
Faculty of Computer & Information Sciences
Hosei University
Tokyo 184-8584, Japan
jianhua@hosei.ac.jp

**ABSTRACT**
There is a growing trend towards the convergence of cyber-physical systems (CPS) and social computing, which will lead to the emergence of smart communities composed of various objects (including both human individuals and physical things) that interact and cooperate with each other. These smart communities promise to enable a number of innovative applications and services that will improve the quality of life. This position paper addresses some opportunities and challenges of building smart communities characterized by cyber-physical and social intelligence.

**INTRODUCTION**
Last decades have witnessed the exponential increase in the number of various computers of everyday use. Modern computers are becoming smaller and smaller, while equipped with higher and higher performance in terms of e.g. computational speed and memory size, as promised by Moore's law. As a consequence, computers are transforming into a lot of new forms. Some examples of new forms of computers include mobile phones, smart sensors, and even ordinary physical things, such as a lamp, a table and a cup. In other words, many physical things will possess computing and communication capabilities of different levels, which are provided by small and (possibly) invisible computers embedded therein. This integration of networked computing and physical dynamics has lead to the emergence of cyber-physical systems (CPS), which have become very hot in recent years.

Generally speaking, CPS features tight integrations of computation, networking, and physical objects, in which various devices are networked to sense, monitor and control the physical world. Although there is no unified definition of this notion, some defining characteristics of CPS include cyber capability in physical objects, networking at multiple and extreme scales, complexity at both temporal and spatial scales, high degrees of automation, and dependable even certifiable operations. Many researchers and practitioners have pointed out that CPS will transform how we interact with the physical world [3,13].

Another recent trend in IT is the rapid growth of social networking services, one of the most impactful concepts in the last decade. Human beings are social animals. Consequently it is unsurprising that social networks have been popular since the beginning of civilization [5]. Over decades they have been playing a significant role in the development of human society. In a social network, a group of individuals are linked through diverse social relationships, such as family, friendships, business partners, and classmates, to mention a few. Such networks have been the subject of research of social scientists for a long time. With the proliferation of Internet technologies such as Web 2.0, online social networking applications have become prevalent recently. Among many examples are Facebook, Twitter, LinkedIn, MySpace, and Bebo. These applications constitute virtual communities that facilitate information creation, distribution, management, sharing and consumption among linked people [1]. For instance, by using social networking services, one can make new friends, stay in touch with (old) friends, and retrieve information of interest anytime and anywhere. Currently most social networking services are web based and provide various means for users to interact over the Internet. Despite the diversity of services provided, a common feature of current social networking applications is that they are designed with the primary goal of facilitating communications among people, that is, we humans are the users.

We envision that the afore-mentioned two trends will converge, at least to some extent, in the foreseeable future. This convergence of CPS and social networking will accelerate the emergence of a new paradigm we call *smart community*. This paradigm will provide us with promising solutions to a variety of societal problems spanning many domains, such as healthcare, education, transportation, energy, and environment, among others. The ultimate goal of building smart communities is to improve the quality of our everyday life by exploiting ubiquitous intelligence [6]. A smart community will by nature feature the interplay of cyber, physical, and social worlds. In this paper, we will illustrate our preliminary vision of smart community as an emerging research area. Some technical challenges facing the field are discussed.

The rest of the paper is organized as follows. In the next section, we first elaborate on the trend towards smart community; then discuss the basic concept of smart community, and identify some of its key characteristics and potential benefits. The third section will outline some key challenges for building smart communities, from a technical perspective. The fourth section concludes the paper.

**VISION OF SMART COMMUNITY**
It is our vision that smart communities built on CPS and social networking services will be ubiquitous. Smart communities will be an integral part of our society of the future, and provide a promising solution to many societal-scale problems. Besides human individuals, social things will become essential elements of smart communities. In a smart community, various social objects, including both humans and physical things, will interact with each other over ubiquitous networks, delivering ubiquitous services by exploring cyber-physical and social intelligence.

**Why Smart Community?**
There are a number of important drivers for building smart communities, among which we would like to mention the following three.

- Our society is facing many critical challenges that remain to be addressed. One of the major challenges is the unprecedented population ageing with the inevitable implications related to disability and care issues. The world's population is ageing rapidly. People are living longer and, as they get older, they are increasingly living alone, and with disabilities. Without modest assistance, the aged, disabled or chronically ill people often live with significantly degraded life quality. Key societal challenges can also be found in other areas such as transportation, energy, environment, and security. In transportation, for example, the traffic networks of many modern cities are getting saturated and congestion becomes endemic. Smart community will offer a constitutive technology to create innovative products, services, and strategies for addressing these societal challenges.

- Smart communities have great potential to improve the quality of life. The design and deployment of smart communities will facilitate constituting the infrastructure needed to create a set of new services for daily life. In assisted living for the elderly, disabled, and chronically ill, for instance, smart communities will make possible, among others, continuous remote monitoring, acute tracking of human motion, real-time emergency response, and remote emergency care [8]. These services will support these impaired people to live more healthy lives, minimizing time in hospital, at local doctors, or in care homes. The ageing population will be able to live more independently in their own homes despite illness or disability, as most people prefer, overcoming isolation and minimizing their reliance on carers. In addition to the domestic environment, ubiquitous services enabled by smart communities can be supplied in, e.g., cars, offices, schools, hospitals, etc. Not only impaired people can benefit from smart communities, but also by healthy people who want to improve the quality of life as well.

- Recent technological advances in sensing, networking, computing are leading to the integration of (wireless) sensor/actuator networks and social networks [1, 2]. By exploring this integration, smart community goes beyond either of the currently isolated two areas, i.e., CPS and social computing. Based on the latest progress in underlying technologies for both areas, smart community can enable many novel functionalities, which otherwise are very difficult, if not impossible, to be realized by either CPS or social networking services alone. For instance, smart communities allow for collaborating sharing among social objects, which can increase real-time awareness of different community members about each other, and provide a better understanding of the aggregate behaviors of the community [1,4]. Massively distributed data collection thus becomes feasible, which leads to the so-called crowd sourcing. It is worth noting that integration of sensor and social networks is not new. In recent years, some efforts have been made in this direction, such as Citysense (http://www.citysense.com/), Brightkite (http://brightkite.com/), and the social life networks (SLN) [5].

**What Is Smart Community?**
Before giving the explanation of the concept of smart community, we'd like to begin this subsection with revisiting the general description of community and online community, two notions closely related to smart community.

*Community and Online Community*
Although there are many different forms of communities, they are undoubtedly vital for humans. There would be no life without community. As a general understanding (according to Wikipedia), a community can be described as a group of interacting people. The group often shares some common values, characteristics or interests, and is attributed with social cohesion. Possibly the most common usage of the term community indicates a large group living in close proximity, generally in social units (i.e. neighborhoods) larger than a household. A large city can be understood as a collection of communities.

In contrast to the communities in real life, an online community is a virtual community that exists online. The members are a group of people communicating or interacting with each other by means of information technologies, typically over the Internet, rather than in person. Social networking services consist in one of the most representative types of online communities. In the information era, online communities have become an important form of social interaction between the majority of people. With the help of modern information technologies, online communities offer the opportunity of instant information exchange, potentially crossing geographical boundaries, which is not possible in a real-world community.

*Smart Community*

In the context of our vision, a *smart community* can be roughly understood as a group of connected (social) objects that interact with each other over ubiquitous networks and deliver smart services to possibly all members. The members of a smart community are objects that can be human individuals, as well as physical things such as a table, a watch, a pen, a door and a key. It is also possible that some other living things (besides human beings) might be included, for example, a tree and a dog. In most cases, these objects have implicit links among them. For example, they may jointly strive to reach a common goal, e.g. sensing and controlling of devices.

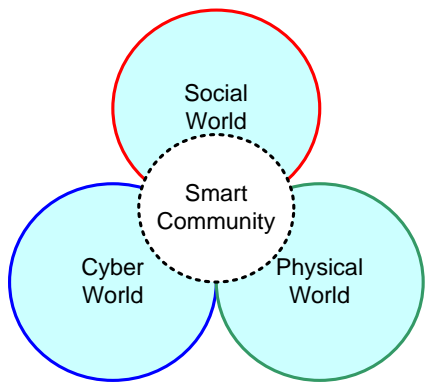

**Figure 1. High-level view of smart community.**

As Fig. 1 illustrates, smart community is an emerging area that features integration and coordination of the cyber, physical, and social worlds. From a convergence point of view, CPS is a result of the integration of the cyberspace and the physical world, while social computing represents the integration of the cyberspace and the human society (i.e. social world). In contrast, smart community, at the intersection of CPS and social computing, indicates a new form of convergence in science and engineering that spans the three major worlds constituting the whole planet. We believe this will cause a paradigm shift in IT.

It is the integration of the cyber, physical, and social worlds that makes it possible for us to built communities that are smart, i.e. capable of behaving in an intelligent manner. From a social perspective, smart communities could enable social awareness among the members by employing certain social sensors [7, 12]. On the other hand, sensor and actuator networks will be exploited to enhance the physical awareness of the context of each object, in e.g. spatial and temporal dimensions. Location data is among the most popular kinds of cyber-physical information to be collected in many applications, thanks to the proliferation of positioning techniques and devices like GPS on mobile phones. When a community of social objects becomes situation aware with respect to both social and cyber-physical contexts, it will be able to observe what is happening, construct a model of its dynamics, conduct spatio-temporal reasoning about future behaviors, and act based on its own decisions [10]. Thus the community will become smart, with the capability of delivering intelligent services to interested users. For example, location-based services could be delivered when the location data is available.

Despite the above observations, it is interesting to make some further comparisons between smart community and CPS, social computing, as well as ubiquitous computing. CPS is about interconnection of (smart) physical things. This is why it is sometimes referred to as internet of things (IoT). Social computing deals with the (virtual) interconnection of people. What a smart community must handle, however, is the hybrid group of people and physical things. Ubiquitous computing, as envisioned by Mark Weiser, is the enabling technology for smart spaces (or smart environments) [6]. In a smart space, humans as users are often the consumers of services, while physical things in the form of various devices act as the service provider. In a smart community we envision, human individuals could be service consumers and/or providers. It is also possible for a physical thing to be a service consumer (in addition to being a service provider).

To this end, several key properties of smart community could be identified as follows.

- A smart community consists of both human individuals and smart physical things as members that interact with each other.

- The scales of smart communities vary case by case. One smart community may have only several members, while another may have a very huge number of members.

- Smart communities are time-evolving. Thus the scale of a smart community may change over time. Consequently, smart communities should have good scalability.

- Smart communities are socially and physically aware systems.

- It is not necessary for smart communities to be Internet based (i.e. online). Some smart communities may function in (local) environments without connection to the Internet.

- The lifecycle of smart communities could possibly be very long in some cases, while relatively short in others, depending on the application supported.

As mentioned above, potential application of smart communities span a variety of domains in which our lives may involve, including e.g. healthcare, education, transportation, energy, environment, and security. One example is collaborative rehabilitation realized by means of integrating body sensor networks and social networks [8]. Another possibility made by smart communities is the construction of global (collaborative) learning environments [9]. Such environments will accommodate physical, social, as well as cyber elements. They offer new opportunities for (distance) education of underserved students by providing resources and expertise that cannot be easily carried through cyber-only methods. Yes another example is a smart transportation community that promises to increase the transportation efficiency of people as well as the whole society.

**TECHNICAL CHALLENGES**

A lot of challenges must be addressed before the full potentials of our vision of smart communities can be explored. In the following, we examine some of these challenges from a technical point of view.

- *Ubiquitous sensing.* To gain a full knowledge of the current situations in different aspects, smart communities demand distributed multimodal data collection in real time. The data collected must be complete and accurate. For this purpose, the sensors deployed, possibly of a very huge number, must perform the sensing tasks in a collaborative way. In addition to physical variables, the social states of community members need to be perceived via social sensing. In this regard, design of social sensors becomes an important issue. Another major challenge is to protect the privacy of individuals without compromising the cyber-physical and social awareness of the system.

- *Autonomous networking.* Network quality of service (QoS) is a common concern of networked systems [13]. Smart communities must support effective and efficient communications between various entities from different worlds (i.e. the cyber, physical, and social worlds), creating a dynamic network of networks. The ability to communicate various types of information across the cyber, physical, and social worlds become necessary. The resulting information flows should be reliable, resource efficient, and in real time, delivering satisfactory QoS. To achieve this goal, autonomous networking and communication protocols need to be developed. In addition, autonomous communications between different communities should be supported.

- *Modeling.* Smart community is a system of systems featuring a high level of complexity. The co-existence of cyber, physical, and social elements within one system causes significant challenges for modeling of community members as well as the community as a whole. For example, does there exist a unified technique for modeling all the entities and the corresponding relationships and interactions? What is the appropriate level of abstraction? Some advanced techniques, such as semantic web, hidden markov models, agent-based stochastic analysis and link-based graph stream analysis, might be employed for community modeling. An issue of particular interest is the modeling of human individuals [11]. It is now possible to study human behaviors through mobile sensing using e.g. (socially aware) smart phones [4, 14]. But building a comprehensive description for every person remains an open issue.

- *Collaborative reasoning.* The capability of reasoning is essential to the intelligence of smart community. Large volumes of data can be collected in smart communities, forming digital footprints of both people and physical objects. In these cases, a critical issue is how to deduce high-level useful knowledge from a large volume of low-level data of disparate types [14]. This problem is challenging because of the inherent heterogeneity of (raw) data. Information distillation and reality mining services can be developed for this purpose. In this respect, it is also important to devise distributed and efficient algorithms for group decision making.

- *Community design and management.* The design and management of smart communities are an open topic yet to be explored. Innovative design principles are needed to cope with the unprecedented complexity of smart communities. A scalable design framework that facilitates specification of a smart community and collaborative design across disciplines is still missing [9]. What are good metrics for measuring the performance of a smart community? How to validate and evaluate the performance? The identification of diverse communities and community members is yet another open issue of great importance. Further research is needed in development of new technologies that support the whole lifecycle of a smart community.

**SUMMARY**

In this position paper, we have described our vision of smart community. A paradigm shift in IT is envisaged with this vision. The cyber, physical, and social worlds will converge in this context. Smart communities will be built upon recent technological advances in both fields of CPS and social computing. We argue that the importance of the role that smart communities will play in our society could not be overly emphasized. In particular, smart communities will help tackle emerging societal challenges in ways that were impossible a few years ago. However, multidisciplinary study must be conducted to address the challenges for building smart communities. We look forward to extensive efforts in this direction.

**ACKNOWLEDGMENTS**

This work was partially supported by the Natural Science Foundation of China under Grant No. 60903153, the Fundamental Research Funds for Central Universities (DUT10ZD110), and the SRF for ROCS, SEM.